\documentclass[useAMS,usenatbib]{mn2e}

\usepackage{graphicx}

\title[A cosmological `probability event horizon' and its observational
implications]{A cosmological `probability event horizon' and its
observational implications}
\author[D. M. Coward and R. R. Burman]{D. M. Coward\thanks{E-mail:
coward@physics.uwa.edu.au}  and R. R. Burman \\School of Physics,
University of Western Australia, M013, Crawley WA 6009, Australia}
\begin{document}

\date{\today}

\pagerange{\pageref{firstpage}--\pageref{lastpage}} \pubyear{2004}

\maketitle

\label{firstpage}

\begin{abstract}
Suppose an astronomer is equipped with a device capable of
detecting emissions --- whether they be electromagnetic,
gravitational, or neutrino --- from transient sources distributed
throughout the cosmos. Because of source rate density evolution
and variation of cosmological volume elements, the sources first
detected when the machine is switched on are likely to be ones in
the high-redshift universe; as observation time increases, rarer,
more local, events will be found. We characterize the observer's
evolving record of events in terms of a `probability event
horizon', converging on the observer from great distances at
enormous speed, and illustrate it by simulating neutron star birth
events distributed throughout the cosmos. As an initial
application of the concept, we determine the approach of this
horizon for gamma-ray bursts (GRBs) by fitting to redshift data.
The event rates required to fit the model are consistent with the
proposed link between core-collapse supernovae and a largely
undetected population of faint GRBs.
\end{abstract}


\section{Introduction}

During the last decade, space-based and terrestrial astronomy have
revolutionized our understanding of the high-$z$ Universe
(redshift $z>1$). Astronomers can now observe the high-$z$
Universe in many bands of the electromagnetic spectrum and have
begun putting upper limits on the gravitational wave spectrum.
Their telescopes and detectors have provided a direct probe of
cataclysmic transient astrophysical phenomena that are rare in our
local Universe but occur at high rates on a cosmological scale.
For example, optical observations of Type Ia supernovae (SNe), are
used as tools to constrain the cosmological model that best
describes our Universe. Their uniformity in luminosity means they
can be used as standard candles for determining cosmological
distances. Recently 23 high-$z$ Type Ia SNe, including 15 at
$z>0.7$ were used as a cosmic ruler to show that the Universe is
accelerating and has cosmological parameters consistent with
spatial flatness \citep{Barris04}.

The high-energy signature of one class of event occurs in the
gamma-ray region of the electromagnetic spectrum: Cosmological
gamma-ray bursts (GRBs) are extremely energetic transient events
and their fluence (time-integrated flux) and time variability show
that their power sources must be compact. At least a fraction of
GRBs are associated with massive and short-lived progenitor stars,
implying that either a neutron star or black hole must be driving
the emissions. Recent estimates suggest that the universal
cosmological GRB rate could be as high as one per minute
\citep{Putten03}.

We adopt these definitions: an {\it event} is a cataclysmic
astrophysical transient occurrence, such as a core-collapse SN or
a GRB, with a duration much less than the observation period; an
{\it observer} records event times and redshifts, obtained as data
from a flux-limited detector sensitive to high-$z$ events.

With the launch of the {\it Swift} satellite (on 2004 November
20), a multi-wavelength GRB observatory
(http://swift.gsfc.nasa.gov/), up to 1000 GRBs are expected to be
detected in 3 years as a result of a sensitivity limit 5 times
fainter than that of BATSE (the Burst And Transient Source
Experiment that was aboard the Compton Gamma Ray Observatory). The
{\it Swift} mission has the potential to localize about one GRB
per day, providing the possibility of cataloging hundreds of GRB
redshifts.

Predicting the rates of transient astrophysical events, such as
GRBs, is fundamental to detector design and planning realistic
science goals. An understanding of the probability of achieving
those goals is correlated with the probability of acquiring a
useful sample of events. The `ultimate detector' would be
sensitive enough to detect all events out to the redshift where
the events of the relevent type first started. For example,
determining the event rate for GRBs as a function of $z$ would
ideally require detectors with all-sky coverage and extraordinary
sensitivity. In addition, selection effects plague many high-$z$
observations.

With continuing technical improvements to astronomical
instrumentation, it is conceivable that the observed rates of GRBs
will approach the theoretical limit in the future. But for the
present, all astronomical detectors probe only a fraction of the
total volume encompassing all possible events. For any
astronomical detector and source type, one can define a
`detectability horizon' centred on the detector and encompassing
the volume in which such events are potentially detectable; the
horizon distance is determined by the flux limit of the detector
and the source flux. A second horizon, defined by the minimum
distance for at least one event to occur over some observation
time, with probability above some selected threshold, can also be
defined. We call this the `probability event horizon' (PEH). It
describes how an observer and all potentially detectable
cosmological events of a particular type are related via a
probability event distribution encompassing all such events.

\section{Probability event horizon}

\subsection{\label{sec:level3}PEH in a Euclidean universe}

To illustrate the idea, we first show how a PEH can be defined in
the simple case of a static Euclidean universe, assuming an
isotropic and homogenous distribution of events. For a constant
event rate $r_0$ per unit volume, the mean cumulative event rate
in a volume of radius $r$ is $ R(r)=(4/3)\pi r^{3}r_0$. The events
are independent of each other, so their distribution is a Poisson
process in time: the probability for at least one event to occur
in this volume during observation time $T$ at a mean rate $R(r)$
at constant probability $\epsilon$ is given by an exponential
distribution:
\begin{equation}\label{prob1}
p(n\ge1;R(r),T) = 1 - e^{-R(r) T}= \epsilon\,,
\end{equation}
$e^{-RT}$ being the probability of zero events occurring. For Eq.
(\ref{prob1}) to remain satisfied as $T$ increases, the mean
number of events in this volume, $N_{\epsilon}= R(r)T=\vert
\mathrm{ln}(1-\epsilon)\vert$, must remain constant. The
corresponding radial distance, a decreasing function of
observation time, defines the PEH:
\begin{equation}\label{horizon}
r_{\epsilon}^{\mathrm{PEH}}(T)= (3N_{\epsilon}/4\pi r_0)^{1/3}
T^{-1/3}\;.
\end{equation}
The speed at which the horizon approaches the observer---the PEH
velocity---is obtained by differentiating $r_{\epsilon}(T)$ with
respect to $T$:
\begin{equation}\label{velocity}
v_{\epsilon}^{\mathrm{PEH}}(T)=(N_{\epsilon}/36 \pi r_0)
^{1/3}T^{-4/3}\,.
\end{equation}

In practice, we'll take $\epsilon=0.95$, corresponding to a 95\%
probability of observing at least one event within distance $r$,
and so $N_{\epsilon}=3$ is the mean number of events. Note that
$r_{\epsilon}^{\mathrm{PEH}}$ and $v_{\epsilon}^{\mathrm{PEH}}$
depend on the choice of $\epsilon$ through $N_{\epsilon}^{1/3}$:
if $\epsilon= 0.1$ is chosen, $N_{\epsilon}=0.1$ and
$r_{\epsilon}^{\mathrm{PEH}}$ and $v_{\epsilon}^{\mathrm{PEH}}$
will be bigger by a factor of 3 than for $\epsilon= 0.95$.

If we consider event rates in volumes of cosmological scale,
$v_{\epsilon}^{\mathrm{PEH}}(T)$ can be very much greater than the
speed of light. For neutron star formation, $r{_0} \approx 5
\times 10^{-12}$ s$^{-1}$ Mpc$^{-3}$ in the nearby Universe
(Coward, Burman \& Blair 2001, section 2.1). Taking this value and
solving Eq. (\ref{velocity}) for $T$ with
$v_{\epsilon}^{\mathrm{PEH}}(T)=c=9.69\times10^{-15}$ Mpc s$^{-1}$
and $\epsilon=0.95$, we find that about $2\times 10^5$ years of
observation time elapse before $v_{\epsilon}^{\mathrm{PEH}}(T)$
drops below $c$. The horizon would then be about 0.2 Mpc from the
observer, but the assumption of homogeneity breaks down for
distances less than a few tens of Mpc. After one year of
observation, the PEH would be 30 Mpc from the observer, which is
about the distance within which we would expect fluctuations in
$R(r)$ resulting from the local distribution of galaxies to
dominate; $v_{\epsilon}^{\mathrm{PEH}}(T)$ at this distance is
about 2 pc s$^{-1}$.

\subsection{Cosmological factors}
To extend the PEH model to a Friedmann cosmology requires
modifying the simple Euclidean model to incorporate cosmic
evolution. The expansion of the Universe is described by the
evolving Hubble parameter $H(z)$, which can be expressed in terms
of the contributions of matter and vacuum energy
\citep{peebles93}:
\begin{equation}\label{hz}
h(z)\equiv H(z)/H_0 = \big[\Omega_{\mathrm m} (1+z)^3+
\Omega_{\mathrm \Lambda} \big]^{1/2}
\end{equation}
for a `flat-$\Lambda$' cosmology (a spatially flat cosmology with
cosmological constant).

The comoving radial (line-of-sight) distance is obtained from
$h(z)$ by
\begin{equation}\label{dp1}
D_\mathrm{c}(z)=\frac{c}{H_0}\int_0^z \frac{{\mathrm d}z}{h(z)}\;;
\end{equation}
this corresponds to integrating elements of comoving
distance---ones that remain constant with epoch for neighbouring
objects moving with the Hubble flow---and is the appropriate
distance measure for structures locked to the Hubble flow
\citep{Hogg}.

The luminosity distance, $D_{\mathrm L}$: defined so that the
inverse-square law of intensity applies, it can be measured by
means of `standard candles'. These two distance measures are
related through redshift by $D_{\mathrm L}=(1+z)D_{\mathrm c}$.

Volume elements in a Friedmann cosmology can be expressed in terms
of $D_{\mathrm c}(z)$ and $H(z)$ (e.g. Porciani \& Madau 2001):
\begin{equation}\label{dvdz}
\renewcommand{\theequation}{S2}
\frac{\mathrm{d}V}{\mathrm{d}z} = 4\pi \frac{c}{H_0}
\frac{{{D_\mathrm{c}}^2(z)}}{h(z)} \;.
\end{equation}

\begin{figure}
\includegraphics[scale=0.7]{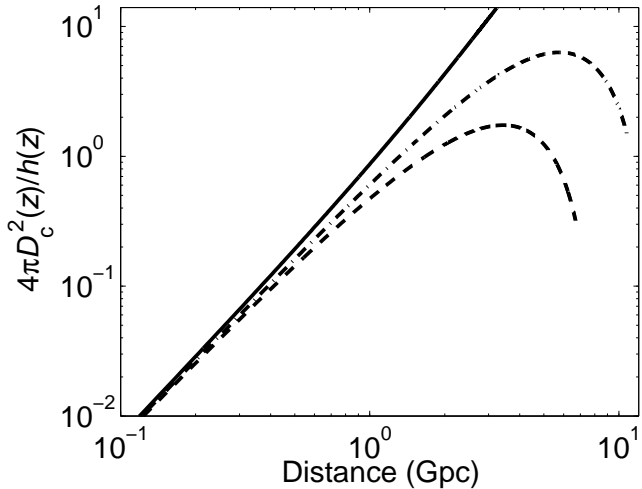}
\caption{The dimensionless volume element per unit redshift,
$\mathrm{d}V/\mathrm{d}z$ (cf Eq. (6) with $c=1=H_0$) as a
function of comoving radial distance for the Euclidean,
flat-$\Lambda$ (0.3, 0.7), and EdS cosmologies; $h(z)\equiv 1$ for
the Euclidean case. Here we use shells of distance rather than
shells of $z$ because redshift is inapplicable for the static
Euclidean cosmology. The curves show that event counts in the
flat-$\Lambda$ and EdS cosmologies reach a maximum and then
decline. This is modified by source rate density evolution and,
for event rates, by time dilation from cosmic expansion. For
distances less than about 1 Gpc the curves converge to that of the
Euclidean cosmology. }\label{volume1}
\end{figure}
In a flat-$\Lambda$ cosmology, the density parameters of matter
and vacuum energy sum to unity; we use $\Omega_{\mathrm m}=0.3$
and $\Omega_{\mathrm \Lambda}=0.7$ for their present-epoch values
and take \mbox{$H_{0}=70$ km s$^{-1}$ Mpc$^{-1}$} for the Hubble
parameter at $z=0$. For comparison, we also consider the
Einstein-de Sitter (EdS) cosmology---a flat Universe with zero
vacuum energy---in which $\Omega_{\mathrm m}=1$ and
$\Omega_{\mathrm \Lambda}=0$. Figure \ref{volume1} plots
$\mathrm{d}V/\mathrm{d}z$ as a function of comoving radial
distance. It shows that volume elements (of fixed $\Delta z$) for
the EdS and flat-$\Lambda$ cosmologies reach a maximum and then
decline, a result of cosmic expansion.

\subsection{Source rate density evolution}
Neutron star (NS) births result from massive short-lived
progenitor stars and there is mounting evidence that GRBs do so
too, so both types of events should closely track the evolving
star formation rate (SFR).

We use a dimensionless SFR density evolution factor $e(z)$,
normalized to unity in our local intergalactic neighborhood, to
account for source rate evolution:
\begin{equation}\label{ez1}
\renewcommand{\theequation}{S4}
e(z) = \frac{(1+W)e^{Qz}}{e^{Rz}+W} \;.
\end{equation}
The two SFR models used in this paper have the parameter values
\begin{enumerate}
\item SF1: $ (Q,R,W)=(3.4, 3.8, 45)$ \item SF2: $(Q,R,W)=(3.4,
3.4, 22)$
\end{enumerate}
as fitted using the EdS cosmology with $H_0=65$ km s$^{-1}$
Mpc$^{-1}$ by \cite{Porc01}. The first star formation history
matches most H$\alpha$ luminosity densities measured in the UV,
and includes an upward correction for dust reddening. The second
model includes more substantial dust extinction at high $z$.

\begin{figure}
\includegraphics[scale=0.65]{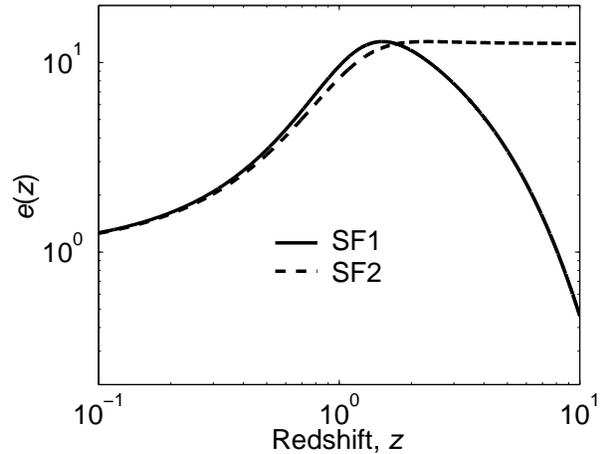}
\caption{Dimensionless evolution factors, $e(z)$, normalized to
unity at $z=0$, based on the parametrized star formation rate
scenarios labelled as SF1 and SF2 in \citet{Porc01} but scaled to
the flat-$\Lambda$ (0.3, 0.7) cosmology; model SF2 allows for high
dust extinction at high $z$.}\label{sfr1}
\end{figure}

An observation-based SFR density is cosmology-dependent, because
it is deduced from the observed spectral luminosity in set
volumes. The above SFR models can be re-scaled to other
cosmologies, as shown in the appendix of \cite{Porc01}. The
luminosity density, and hence SFR density, are proportional to
${D_{\mathrm{L}}}^2/(\mathrm{d}V/\mathrm{d}z)$, which is
proportional to $(1+z)^2 H(z)$, and so both scale as $H(z)$ among
different cosmologies. But $e(z)$, which is normalized to unity at
$z=0$, just follows the shape of the SFR density curve and so
scales as $h(z)$; that is $e(z)/h(z)$ is invariant under change of
cosmology. For converting the above evolution factors to a
flat-$\Lambda$ cosmology with $\Omega_{\mathrm{m}}=0.3$ and
$\Omega_{\Lambda}=0.7$, the scale factor is
\begin{equation}\label{ez1}
[0.3(1+z)^3+0.7]^{1/2}(1+z)^{-3/2}\,,
\end{equation}
as follows from Eq. (\ref{hz}) for $h(z)$. The resulting $e(z)$
functions are shown in Fig. \ref{sfr1}.

\subsection{PEH in flat-$\Lambda$ and EdS cosmologies}
\begin{figure}
\includegraphics[scale=0.65]{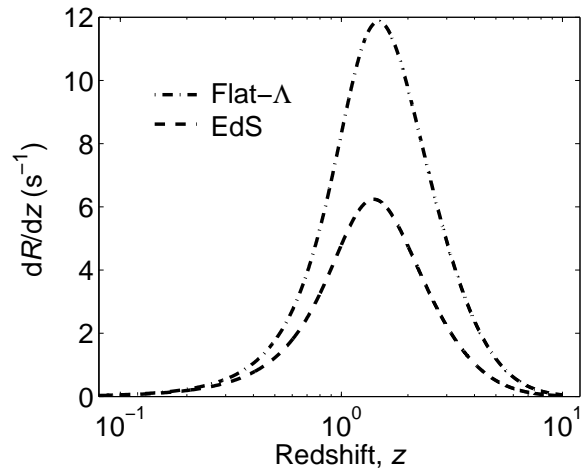}
\caption{The differential neutron star birth rate,
$\mathrm{d}R/\mathrm{d}z$, as a function of redshift, for the
flat-$\Lambda$ (0.3, 0.7) and EdS cosmologies and the star
formation rate model labelled SF1 in \citet{Porc01} but scaled to
the flat-$\Lambda$ (0.3, 0.7) cosmology. The present-epoch rate
$r_0$ is taken to be $5\times 10^{-12}$ s$^{-1}$ Mpc$^{-3}$.}
\label{fdrdz}
\end{figure}
\begin{figure}
\includegraphics[scale=0.65]{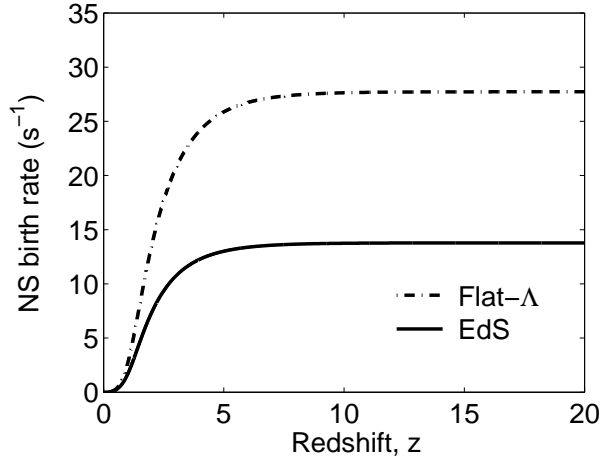}
\caption{The cumulative neutron star birth rate, $R(z)$, obtained
by integrating $\mathrm{d}R/\mathrm{d}z$ from $z=0$. Universal
event rates (integrated throughout the cosmos) are 14 and 27
s$^{-1}$ in the EdS and flat-$\Lambda$ cosmologies respectively.}
\label{rates}
\end{figure}

In the Euclidean model, the event rate in the shell $r$ to
$r+{\mathrm d}r$ centred on the observer is $\mathrm{d}R = 4\pi
r^2 r_0 \mathrm{d}r$. In the standard Friedmann cosmologies, one
can express the differential event rate as the event rate in the
redshift shell $z$ to $z+{\mathrm d}z$:
\begin{equation}\label{drdz}
\mathrm{d}R = \frac{\mathrm{d}V}{\mathrm{d}z}\frac{r_0 e(z)}{1+z}
\mathrm{d}z \,,
\end{equation}
\noindent where $\mathrm{d}V$ is the cosmology-dependent co-moving
volume element and  $R(z)$ is the all-sky ($4\pi$ solid angle)
event rate, as observed in our local frame, for sources out to
redshift $z$. Source rate density evolution is accounted for by
the dimensionless evolution factor $e(z)$; as this is normalized
to unity in the present-epoch universe $(z=0)$, $r_0$ is now the
$z=0$ rate density. The $(1 + z)$ factor accounts for the time
dilation of the observed rate by cosmic expansion, converting a
source-count equation to an event rate equation.

The cumulative NS birth or GRB event rate $R(z)$ is calculated by
integrating the differential rate from the present epoch to
redshift $z$. It depends on cosmology through the factors
$\mathrm{d}V/\mathrm{d}z$ and $e(z)$ in Eq. (\ref{drdz}) for
$\mathrm{d}R$. The differential and cumulative rates of NS
formation as functions of $z$ are plotted in Figs. \ref{fdrdz} and
\ref{rates}, for model SF1. Figure \ref{rates} shows that $R(z)$
is higher in the flat-$\Lambda$ (0.3, 0.7) cosmology than in the
EdS one. Because the curves scale with $r_0$, they can also be
applied to GRBs, but with the rates a factor of about $10^6$ lower
for the classical GRBs.

For event types locked to the star formation rate, $e(z)$ is
obtained by normalizing a SFR model to its $z=0$ value. Presently
there is no consensus among astronomers on a model of star
formation history at high $z$, so we shall use either a constant
star formation history or the parametrized models of
\citet{Porc01}.

Integrating the differential rate yields the cosmological analogue
of the Euclidean mean cumulative event rate:
\begin{equation}
R(z)=\int_{0}^{z}(\mathrm{d}R/\mathrm{d}z)\;\mathrm{d}z\,.
\end{equation}
Because of variation of cosmological volume elements (Fig.1), the
rate derivative $\mathrm{d}R/\mathrm{d}z$ reaches a maximum at
high $z$ in both the flat-$\Lambda$ and EdS cosmologies.
Consequently there is a slower growth in the cumulative rate as
compared to that in the simple Euclidean model in which the volume
elements grow as $r^2$.
\begin{figure}
\includegraphics[scale=0.65]{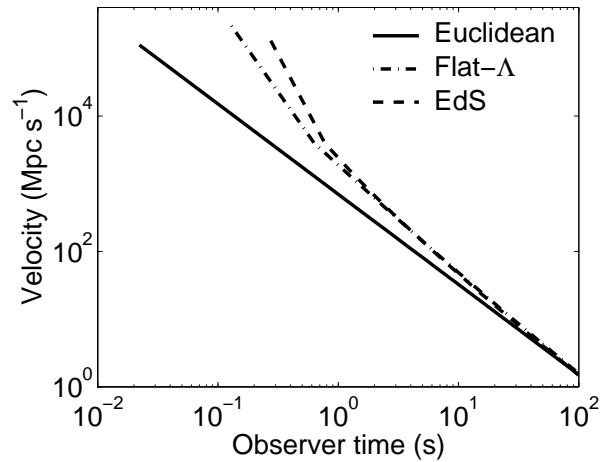}
\caption{The probability event horizon (PEH) velocity for
Euclidean, flat-$\Lambda$ (0.3, 0.7) and EdS cosmologies assuming
a constant event rate density of $5 \times 10^{-12}$ s$^{-1}$
Mpc$^{-3}$. The starting points at the upper left indicate that
the PEH is `stalled' until the number of events exceeds some
threshold, which is cosmology dependent as a result of different
cumulative event rate functions. The flat-$\Lambda$ and EdS curves
depart markedly from the Euclidean curve during the first seconds
of observation time. The change in gradient apparent in the
flat-$\Lambda$ and EdS curves occurs because of a slower increase
in volume at high $z$.}\label{simsv1}
\end{figure}

The events follow the same probability distribution as in the
Euclidean case (Eq. \ref{prob1}) but with $R$ now expressed as a
function of $z$:
\begin{equation}\label{prob2}
p(n\ge1;R(z),T) = 1 - e^{-R(z) T}= \epsilon\,,
\end{equation}
with $N_{\epsilon}\equiv R(z)T=\vert \mathrm{ln}(1-\epsilon)\vert$
constant. Instead of inverting to find
$r_{\epsilon}^{\mathrm{PEH}}(T)$, we find
$z_{\epsilon}^{\mathrm{PEH}}(T)$ by solving this condition
numerically, thus defining the PEH. Converting
$z_{\epsilon}^{\mathrm{PEH}}(T)$ to luminosity distance and
differentiating with respect to $T$ then yields
$v_{\epsilon}^{\mathrm{PEH}}(T)$.

\begin{figure}
\includegraphics[scale=0.65]{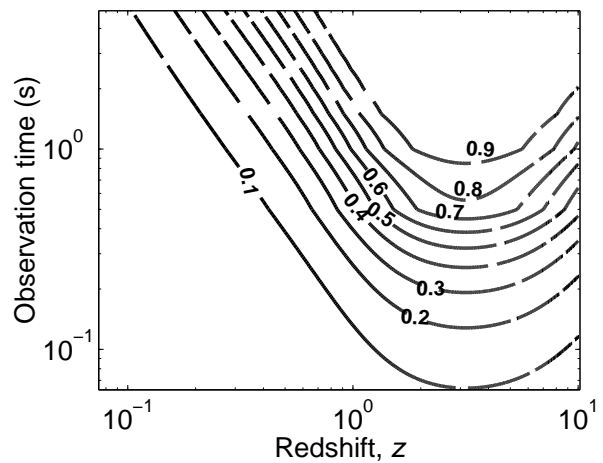}
\caption{Contours, based on the same constant event rate density
and flat-$\Lambda$ cosmology as used in Fig. \ref{simsv1}, show
the probability (labelling the curves) for at least one event to
occur in the redshift shell $z$ to $z+\mathrm{d}z$ as a function
of $z$ and observation time. The minimum of each of the contours
corresponds to the most probable redshift at which an event will
occur. As a function of observation time, each probability contour
follows two trajectories
--- one approaching the observer and the other receding. The receding contours reach the
redshift where events first started within a fraction of a second,
as opposed to years in observer time for an approaching
probability contour to reach the local Universe.} \label{contour}
\end{figure}

\section{Neutron star birth events} Figure \ref{simsv1} shows the PEH
velocity for NS birth events in the three cosmologies ---
Euclidean, flat-$\Lambda$ and EdS---with no source evolution so
$e(z)=1$. Because the local Universe is approximately Euclidean,
the PEH velocity converges to the Euclidean result when the
observation time becomes much longer than the mean time interval
between events. The cumulative rate, $R(z)$, is cosmology
dependent, because the differential rate is determined by the
evolution of volume elements as a function of $z$ (Eq.
\ref{dvdz}). Variations in event rates result in differing PEH
velocities; for example, the relatively smaller event rate
obtained using the EdS model (see Fig. \ref{rates}) manifests as a
longer {\it time delay} before the initiation of PEH motion
because $p(n\ge1;R(z),T)$ remains below 0.95 for a longer
observation time.

Figure \ref{contour} shows probability contours, in the
flat-$\Lambda$ cosmology, for at least one event to occur in the
redshift shell $z$ to $z+{\mathrm d}z$ as a function of $T$ and
$z$. The curves reveal two horizons, one approaching and one
receding as $T$ increases; the receding horizon, corresponding to
the observation of increasingly distant events beyond the peak of
$\mathrm{d}R/\mathrm{d}z$, is very short lived.

\begin{figure}
\includegraphics[scale=0.65]{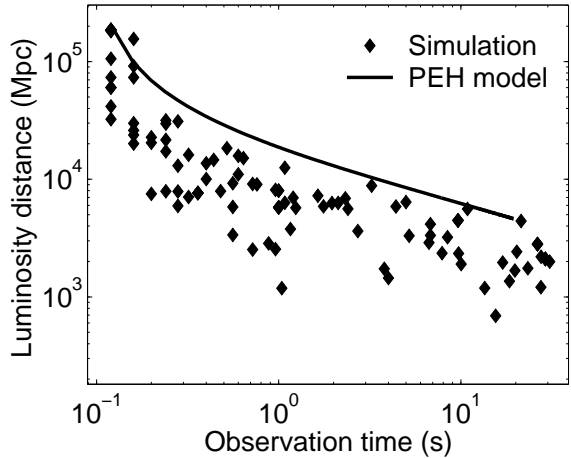}
\caption{Simulation illustrating the rapid motion of the PEH for
the first few tens of seconds of observation time, and a model
curve with a probability threshold $\epsilon=0.95$ for at least
one event to occur at a distance less than that of the PEH. We
assume a universal cumulative event rate of about 25 s$^{-1}$ as
seen in our frame, comparable to the NS birth rate integrated
throughout the Universe, a flat-$\Lambda$ (0.3, 0.7) cosmology and
a source rate evolution locked to the SFR model labelled SF2 in
Porciani \& Madau (2001)} \label{simsv}
\end{figure}

\begin{figure}
\includegraphics[scale=0.65]{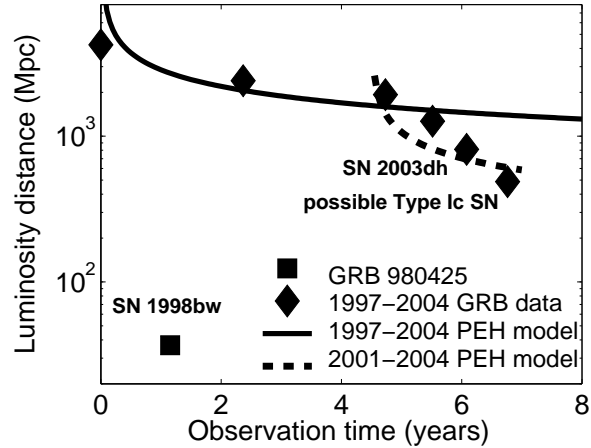}
\caption{Filled diamonds show the closest events as a function of
observation time for the 34 GRBs for which the redshift is known
with reasonable confidence; the solid square is the highly
under-luminous GRB 980425. The observations span a 7-yr period,
starting from the first redshift measurement in 1997 February  and
continuing to 2004 July. The solid curve shows the PEH, excluding
the outlier GRB980425, assuming a flat-$\Lambda$ (0.3, 0.7)
cosmology and a source rate evolution based on the star formation
rate model labelled SF1 in Porciani \& Madau (2001). A local rate
density $r_0\approx 0.8$ yr$^{-1}$ Gpc$^{-3}$ is used to fit to
the redshift data for the classical GRBs for the first 4--5 yr of
observation. (The probability of GRB 980425 occurring inside a
volume bounded by its redshift $z=0.0085$ in this time is
$5\times10^{-5}$.) After this period, a more rapidly decreasing
trend in distance is observed. The dashed curve shows the PEH
model fitted to the data from 2001 September to 2004 July  with
$r_0\approx 27$ yr$^{-1}$ Gpc$^{-3}$. GRBs associated with
core-collapse SNe are labelled.} \label{GRB-data}
\end{figure}

To demonstrate the motion of a PEH we have simulated the
occurrence, as a function of observation time, of NS birth events
distributed throughout the Universe. The redshifts are treated as
a random variable following a probability distribution function
obtained by normalizing $dR/dz$ (Coward, Burman \& Blair 2002a,b):
\begin{equation}
p(z){\mathrm d}z = {\mathrm d}R/R(\infty) \;,
\end{equation}
$R(\infty)$ being the Universal rate, integrated throughout the
cosmos, as seen in our frame. The corresponding cumulative
distribution function $C(z)$, giving the probability of an event
occurring in the redshift range $0$ to $z$, is
\begin{equation}
C(z) = R(z)/R(\infty) \;.
\end{equation}
We use the inverse function of $C(z)$ to produce values of $z$ by
employing a random number generator to select values from $C(z)$.
\par
Our simulation uses $r_0 \approx 5\times10^{-12}$ s$^{-1}$
Mpc$^{-3}$ for the $z=0$ rate density of these events and an
evolution locked to the SFR model labelled SF2 in Porciani \&
Madau (2001) scaled to the flat-$\Lambda$ $(0.3,0.7)$ cosmology.
The running minimum redshifts of the simulated events as a
function of observation time define the PEH; these have been
converted to luminosity distances and plotted in Fig. 7 along with
a PEH curve for $\epsilon = 0.95$. Of these 103 events only one is
at a distance greater than the PEH. The binomial 95\% confidence
interval, give $0.9471<P<0.9998$ for one failure in 103 trials,
which is in good agreement with $\epsilon=0.95$ for at least one
event to occur within the PEH.

In this calculation of the approach of the PEH for simulated NS
birth events throughout the Universe, we have assumed that the
observation time for each event is small compared with the typical
observed temporal separation of events, which averages about 40
ms. The calculation is directed toward future `Advanced LIGO'-type
gravitational wave detectors, which may be able to detect a
stochastic background of gravitational radiation generated by
these events. Recent calculations, based on the computed wave
functions of \cite{Dimm02}, suggest that the duty cycle of the
signal will be well below one (Howell et al., 2004, figure 16),
implying a signal consisting largely of non-overlapping
wave-trains, described as `popcorn noise'.

For optical surveys of very high-$z$ SNe, selection effects and
the sensitivity limits of the detectors reduce the observed rate
substantially. If these effects were well enough understood, then
the rate density of events could still be inferred from the
evolving record of observations. These ideas will be demonstrated
in the next section for GRB afterglow observations.

\begin{table*}
 \centering
 \begin{minipage}{140mm}
\caption{\label{tab:example}GRB redshifts adapted from the table
of localized GRBs at {\tt
http://www.aip.de/{$\sim$}jcg/grbgen.html}. The instruments used
to localize the GRBs are listed along with confirmation if the
Interplanetary Network (IPN) was used. Other columns show: XA =
X-ray afterglow; OT = optical (or IR) transient; RA = radio
afterglow; IAUC (International Astronomical Union Circular)
number; $z$ = redshift. An `X' after the date marks events with no
gamma-ray detection, but with X-ray properties similar to GRBs --
so-called X-ray flashes. The `C' after one of the dates identifies
it as the 3rd GRB on that day. The running minimum observed GRB
redshifts from 1997 February to 2004 July  (entries highlighted in
bold) are used in Fig. \ref{GRB-data}, with GRB 980425 regarded as
an outlier.\vspace{0.2cm}}
\begin{tabular}{|l|l|l|l|l|l|l|l|}\hline
  GRB & Instrument & IPN & XA & OT & RA & IAUC & $z$\\\hline

  {\bf 970228} & SAX/WFC & y & y & y & n & 6572 & {\bf 0.695} \\
  970508 & SAX/WFC &  & y & y & y & 6649, 6654 & 0.835  \\
  970828 & SAX/WFC & y & y & n & y & 6726, 6728 & 0.9578  \\
  971214 & SAX/WFC & y & y & y & n & 6787, 6789 & 3.42  \\
  {\bf 980425} & SAX/WFC &   & y & SN & y & 6884 & {\bf 0.0085} \\
  980613 & SAX/WFC &  & y & y & n & 6938 & 1.096  \\
  980703 & RXTE/ASM &  & y & y & y & 6966 & 0.966  \\
  990123 & SAX/WFC & y & y & y & y & 7095 & 1.6  \\
  990506 & BAT/PCA & y & y &  & y & - & 1.3  \\
  990510 & SAX/WFC & y & y & y & y & 7160 & 1.619  \\
  990705 & SAX/WFC & y & y & y & n & 7218 & 0.86  \\
  {\bf 990712} & SAX/WFC &  &  & y & n & 7221 & {\bf 0.434}  \\
  991208 & Uly/KO/NE & y &  & y & y &  & 0.706  \\
  000131 & Uly/KO/NE & y &  & y &  &  & 4.5  \\
  000210 & SAX/WFC & y & y &  & y &  & 0.846  \\
  000301C & ASM/Uly & y & & y & y &  & 2.03  \\
  000911 & Uly/KO/NE & y & & y & y &  & 1.058  \\
  000926 & Uly/KO/NE & y & y & y & y &  & 2.066  \\
  010921 & HE/Uly/SAX & y &  & y &  &  & 0.45  \\
  {\bf 011121} & SAX/WFC & y & y & y & y &  & {\bf 0.36}  \\
  011211 & SAX/WFC &  & y & y &  &  & 2.14  \\
  020124 & HETE & y &  & y &  &  & 3.198  \\
  020405 & Uly/MO/SAX & y &  & y & y &  & 0.69  \\
  020813 & HETE & y & y & y & y &  & 1.25  \\
  {\bf 020903X} & HETE &  &  & y & y &  & {\bf 0.25}  \\
  021004 & HETE &  & y & y & y &  & 2.3  \\
  021211 & HETE &  &  & y &  &  & 1.01  \\
  030226 & HETE &  & y & y &  &  & 1.98  \\
  030323 & HETE &  &  &  & y &  & 3.372  \\
  030328 & HETE &  & y & y &  &  & 1.52  \\
  {\bf 030329} & HETE &  & y & y & y & 8101 & {\bf 0.168}  \\
  030429 & HETE &  &  & y &  &  & 2.65  \\
  {\bf 031203} & INTEGRAL &  & y & y & y & 8250, 8308 & {\bf 0.105}  \\
  040701X & HETE &  & y &  &  &  & 0.2146  \\
\hline
\end{tabular}
\end{minipage}
\end{table*}

\section{GRB events}
The PEH velocity is based on the {\it act} of observation, where
the observer is at the centre of a Universe defined in terms of an
event probability distribution. The temporal evolution of the PEH
for a particular class of cosmological transient event is defined
by the history of event rates. Different types of events, for
example NS births and GRBs, will have different PEH velocities
because their mean event rate densities differ, even if both are
locked to the SFR. The PEH concept can be applied to the observed
GRB redshift distribution assuming some knowledge of source-rate
evolution, luminosity distribution of the sources, observational
selection effects and the local rate density; these parameters are
all uncertain.

We use GRB redshift data from 1997 Febuary (when the first GRB
redshift was obtained) to 2004 July, which consist of 235 GRBs
localized to less than 1 degree within a few hours to days. About
40 GRB redshifts have been measured but only 34 are known with
certainty (mostly identified from their host galaxies); these
events, along with details of each observation, are shown in Table
1.

Figure \ref{GRB-data} plots the temporal evolution of the minimum
observed luminosity distance, using these measured GRB redshifts,
as a function of observation time from 1997 February to 2004 July;
we do not include GRB redshifts that are highly uncertain. We note
that the very under-luminous and nearby GRB 980425 ($z=0.0085$
corresponding to 40 Mpc), and GRBs 030329 $(z=0.168)$ and 031203
$(z=0.105)$, are associated with cc SNe, probably Type Ib/c.

Our data points are obtained by selecting the sources with minimum
redshift as a function of observation time over the GRB redshift
data set. A PEH model curve is fitted to these data, assuming the
flat-$\Lambda$ (0.3, 0.7) cosmology and a source rate evolution
locked to the SFR model labelled SF1 in \cite{Porc01}. The
differential rate has one free parameter, the present-epoch GRB
rate density, $r_0$, which we vary in the PEH model to fit the
data. We scale the differential event rate by a GRB {\it
luminosity function} following a log-normal distribution
\citep{bromm02}, with a mean GRB luminosity of $2\times10^{56}$
photons s$^{-1}$, and assume a photon flux threshold of 0.2
s$^{-1}$ cm$^{-2}$ for BATSE (see Appendix A). We note that the
form of luminosity distribution that would fit both the
`classical' GRBs and GRBs 980425 and 031203 is not apparent.

Excluding the outlier GRB 980425, we find that the data for the
first 4--5 yr of observation time (classical GRBs) are best fitted
using a PEH curve with $r_0\approx 6\times 10^{-2}$ yr$^{-1}$
Gpc$^{-3}$. Scaling this rate by accounting for the solid angle
monitored by the BeppoSAX camera or HETE telescope, about 0.9 str
\citep{Guetta04}, yields $r_0\approx 0.8$ yr$^{-1}$ Gpc$^{-3}$,
similar to the rate density of 0.5 yr$^{-1}$ Gpc$^{-3}$ deduced
from observations of both optical and optically dark GRBs
\citep{schmidt01,Frail}. The probability of at least one GRB
occurring inside a volume bounded by $z=0.17$ (in 4.5 yrs
observation time) has the quite significant value of 0.37, but the
probability of one occurring inside a volume bounded by $z=0.0085$
is only about $5\times10^{-5}$: it is difficult to reconcile the
very nearby GRB 980425 with the redshift distribution of classical
GRBs. \citet{soderberg04} conclude that the two under-luminous
GRBs 980425 and 031203 were intrinsically sub-energetic and
un-jetted events.

Data from 2001--2004, which include GRBs 030329 and 031203, show a
more sharply decreasing trend in Fig. \ref{GRB-data}, implying a
relatively higher PEH velocity and so a higher $R(z)$ and hence a
higher $r_0$; fitting a PEH curve to this subset of data requires
$r_0\approx 27$ yr$^{-1}$ Gpc$^{-3}$. With this $r_0$, the
probability for an event to occur in the volume bounded by GRB
980425 $(z=0.0085)$ in 4.5 yr is 0.0016. Thus, it seems likely
that GRB 980425 is a member of a more numerous population of
low-luminosity GRBs.

Because GRBs 980425 and 031203 were sub-energetic and showed no
evidence for jetted emissions, $r_0$ for such a low-luminosity GRB
population could be orders of magnitude higher than that of the
classical GRB population, but still only a small fraction of the
Type Ib/c SN rate \citep{cow04}.

\section{Concluding remarks}
The probability event horizon concept provides a means of
analyzing an observer's evolving record of events in terms of the
distribution of the sources throughout the cosmos. It is
particularly sensitive to the shape of the low probability `tail'
of the source distribution. The dependence of the PEH velocity on
the local source rate density means that by analyzing the redshift
distribution of the sources as a function of observation time one
can infer the local source rate density.

As an initial application of this, we have shown how fitting a PEH
model to GRB redshift data can yield information on the local GRB
rate density. Such an analysis can be used to probe the progenitor
populations that comprise the distributions. This is demonstrated
by Figure \ref{GRB-data}, where GRB 980425 is either a statistical
outlier of the classical population or a member of a different
distribution of GRBs that could be intrinsically faint;
alternatively, it could be a classical GRB observed away from its
jet axis.

The PEH concept has applications to gravitational wave detection.
Understanding the detectability of astrophysical sources as a
function observation time is crucial to these searches. Current
detectors are only sensitive to local events, such as binary
coalescences out to several tens of Mpc; the PEH for these events
corresponds to long observation times of the order of a year or
so. Because of the rarity of such events, we are modelling the
detectability of more frequent but fainter gravitational wave
sources using the PEH concept.

We are currently studying the application of the PEH in
simulations of the evolving spectra of contributions to the
stochastic gravitational wave background generated by NS births
and other cataclysmic events distributed throughout the Universe.

\section*{Acknowledgments}
We thank the referee for helpful comments that have clarified
several apects of this paper. We thank D. G. Blair and E. J.
Howell for discussions, and B. P. Schmidt and M. H. P. M. van
Putten for reading and commenting on early versions of this work.
D. M. Coward is supported by an Australian Research Council
fellowship and grant DP0346344.

\appendix

\section[]{GRB Luminosity function}
The GRB luminosity function (LF), together with the flux
sensitivity threshold of the instrument, determines the fraction
of all GRBs potentially detectable with that instrument
\citep{norris02}. Following \cite{bromm02}, we scale
$\mathrm{d}R/\mathrm{d}z$ to account for the distribution of GRB
luminosities:
\begin{equation}\label{grbrate}
\psi_{\rm GRB}(z)= \int_{L_{\rm lim}(z)}^{\infty}p(L) {\rm d}L \;,
\end{equation}
where $\psi_{\rm GRB}(z)$ is the GRB rate scaling function and
$p(L)$ is the GRB LF with $L$ the intrinsic luminosity in units of
photons s$^{-1}$. With $f_{\rm lim}$ denoting the flux sensitivity
threshold, in photons s$^{-1}$ m$^{-2}$, the minimum detectable
luminosity can be expressed as a function of redshift by $L_{\rm
lim}(z)= 4 \pi {D_{\mathrm L}}^{2}(z) f_{\rm lim}$, with
$D_{\mathrm L}(z)$ the luminosity distance.

In modelling the GRB redshift distribution we use the log-normal
LF cited by \cite{bromm02}:
\begin{equation}\label{PL}
p(L)= \frac{{\rm e}^{-\sigma^{2}/2}}{\sqrt{2\pi \sigma^{2}}} \exp
\bigg\{-\frac{[\ln(L/L_{0})]^{2}}{2\sigma^{2}}\bigg\}\frac{1}{L_{0}}
\;,
\end{equation}
where $\sigma$ and $L_{0}$ are the width and average luminosity,
respectively. We assume $\sigma=2$ and $L_{0}=2\times 10^{56}$
s$^{-1}$, with $f_{\rm lim}=0.2$ photons s$^{-1}$ cm$^{-2}$ for
BATSE \citep{bromm02}.

\bsp \label{lastpage}


\begin{thebibliography}{99}

\bibitem[\protect\citeauthoryear{Barris et al.}{2004}]{Barris04} Barris B.J. et al., 2004, ApJ, 602, 571
\bibitem[\protect\citeauthoryear{Bromm \& Loeb}{2002}]{bromm02}Bromm V., Loeb A., 2002, ApJ, 575, 111
\bibitem[\protect\citeauthoryear{Coward, Burman \& Blair}{2001}]{cow01}Coward D.M., Burman R.R., Blair D.G., 2001, MNRAS, 324, 1015
\bibitem[\protect\citeauthoryear{Coward, Burman \& Blair}{2002}]{cow02a}Coward D.M., Burman R.R., Blair D.G., 2002a,
MNRAS, Soc. 329, 411
\bibitem[\protect\citeauthoryear{Coward, Burman \& Blair}{2002}]{cow02b}Coward D., Burman R., Blair D., 2002b, Class. Quantum Grav.
19, 1303
\bibitem[\protect\citeauthoryear{Coward}{2005}]{cow04} Coward D.M., MNRAS, 2005 (submitted)
\bibitem[\protect\citeauthoryear{Dimmelmeier, Font \& M\"uller}{2002}]{Dimm02} Dimmelmeier H., Font, J.A., M\"uller, E., 2002, A\&A,
393, 523
\bibitem[\protect\citeauthoryear{Frail et al.}{2001}]{Frail}Frail D.A. et al., 2001, ApJ, 562, L55
\bibitem[\protect\citeauthoryear{Guetta}{2004}]{Guetta04} Guetta D., Perna R., Stella L., Vietri M., 2004, ApJ, 615, L73
\bibitem[\protect\citeauthoryear{Hogg}{2000}]{Hogg}Hogg D.W., 2000, astro-ph/9905116 v4
\item Howell E., Coward D., Burman R., Blair D., MNRAS, 2004, 351,
1237
\bibitem[\protect\citeauthoryear{Norris}{2002}]{norris02} Norris J.P., 2002, ApJ, 579, 386
\bibitem[\protect\citeauthoryear{Peebles}{1993}]{peebles93}Peebles P.J.E., 1993, Principles of Physical Cosmology, Princeton Univ. Press, Princeton, New Jersey, p. 312
\bibitem[\protect\citeauthoryear{Porciani \& Madau}{2001}]{Porc01}Porciani C., Madau P., 2001, ApJ., 548, 522
\bibitem[\protect\citeauthoryear{Schmidt}{2001}]{schmidt01} Schmidt M., 2001, ApJ, 552, 36
\bibitem[\protect\citeauthoryear{Soderberg et al.}{2004}]{soderberg04} Soderberg A.M. et al., 2004, Nat., 430, 648
\bibitem[\protect\citeauthoryear{van Putten \& Regimbau}{2004}]{Putten03} van Putten M.H.P.M., Regimbau T., 2003, ApJ, 593, L15
\end{thebibliography}
\end{document}